# Cluster dynamics modeling of Mn-Ni-Si precipitates in ferritic-martensitic steel under irradiation

Jia-Hong Ke [a], Huibin Ke [a], G. Robert Odette [b] and Dane Morgan [a*]

[a] Department of Materials Science and Engineering, University of Wisconsin-Madison, Madison, WI 53706, USA

[b] Materials Department, University of California, Santa Barbara, CA 93106, USA

* Corresponding author. E-mail address: ddmorgan@wisc.edu

## Abstract

Mn-Ni-Si precipitates (MNSPs) are known to be responsible for irradiation-induced hardening and embrittlement in structural alloys used in nuclear reactors. Studies have shown that precipitation of the MNSPs in 9-Cr ferritic-martensitic (F-M) alloys, such as T91, is strongly associated with heterogeneous nucleation on dislocations, coupled with radiation-induced solute segregation to these sinks. Therefore it is important to develop advanced predictive models for Mn-Ni-Si precipitation in F-M alloys under irradiation based on an understanding of the underlying mechanisms. Here we use a cluster dynamics model, which includes multiple effects of dislocations, to study the evolution of MNSPs in a commercial F-M alloy T91. The model predictions are calibrated by data from proton irradiation experiments at 400 °C. Radiation induced solute segregation at dislocations is evaluated by a continuum model that is integrated into the cluster dynamics simulations, including the effects of dislocations as heterogeneous nucleation sites. The result shows that MNSPs in T91 are primarily irradiation-induced and, in particular, both heterogeneous nucleation and radiation-induced segregation at dislocations are necessary to rationalize the experimental observations.

# 1 Introduction

The extremely harsh environment in advanced fission and fusion reactors requires the development of structural materials with superior performance and stability. Ferritic-martensitic (F-M) steels are a leading candidate structural alloy because of their attractive and well-established properties including reduced activation, swelling resistance and irradiation stability [1]. Nevertheless, irradiation drives microstructural, microchemical and precipitate evolutions in F-M alloys [2-5], which can degrade the performance and safe lifetime limits of advanced reactor structural components. For example, radiation-induced precipitates cause hardening and embrittlement by acting as obstacles to dislocation glide. Thus understanding and predicting precipitate development as a function of the alloy composition and starting microstructure, as well as the irradiation conditions, are critical to a wide range of nuclear energy technologies.

In this work we focus on the alloy T91, a commercial F-M steel with ~9% Cr and small additions of other micro-alloying elements such as C, Ni, Mn, Si, V, Mo, etc. Special attention is given to the precipitation of MNSPs that have been shown to cause high fluence embrittlement of reactor pressure vessel (RPV) steels [6-9]. Radiation-induced segregation (RIS), solute clustering and precipitation of minor elements in T91 have been investigated experimentally in a number of recent studies. Jiao and Was [3] and Wharry *et al.* [4] reported segregation of Cr, Ni, and Si at defect sinks such as dislocation lines/loops, grain boundaries and precipitate/matrix interfaces in T91 under proton and heavy-ion irradiation. Wharry and Was [10] studied the temperature dependence of RIS at grain boundaries and found that enrichment of minor element (Ni, Si, Cu) RIS peaked at 400-500 °C. Both MNSPs and Cu-rich precipitates (CRPs) have been observed by atom probe tomography (APT) following proton-irradiated T91 at 400 and 500 °C [4]. However, Jiao *et al.* [11] noted that no MNSPs were observed in T91 irradiated at less than 1 dpa (displacements per atom) at 400 °C under proton irradiation at a dose rate of $10^{-5}$ dpa/s, suggesting significant kinetic limits on their formation. In both MNSP studies in T91 [4, 11], dislocations were found to be the preferred MNSP nucleation site. Wharry *et al*. [4] also suggested that the strong segregation of Si at dislocations due to RIS may correlate with the evolution of the MNSPs.

Simulation studies of MNSPs in steels have been carried out by calculation of phase diagrams (CALPHAD) methods [12], lattice Monte Carlo simulations [13, 14] and very recently by cluster dynamics (CD) modeling [7, 15]. A recent Monte Carlo study showed that dislocation loops can act as heterogeneous nucleation sites for solute clusters in RPV steels [16]. Solute clusters with enriched Ni, Si, P and Cr were found in irradiated Fe-Cr model alloys [17-20]. However, previous modeling studies have not focused on MNSPs in F-M alloys, including T91. Additionally, RIS and dislocation effects on the density and size evolution of MNSPs are not well understood. Here we utilize the CD model to predict the evolution of MNSPs in a commercial F-M alloy T91 and compare with available experimental data. The model considers multiple effects of dislocation, including heterogeneous nucleation and radiation-induced solute segregation at dislocations.

## 2  Methods

We employ thermodynamic analysis and CD modeling [21] to study the stability and evolution of MNSPs in T91. The CD method has been utilized successfully to study nucleation-growth processes of diffusional phase transformations, and recent applications include Cu precipitation in α-Fe [22], oxide precipitation in nanostructured ferritic alloys [23], and MNSPs in RPV steels under irradiation [7, 15]. Governing equations and parameters used in the CD model are summarized in the Supplementary Information (SI Section S1). The CD simulations includes an embedded radiation-enhanced diffusion (RED) model developed by Odette *et al*. [24]. The thermodynamic driving force of MNSP nucleation is calculated based on the TCAL3 database. To account for the multiple effects of dislocations, we combine the CD model with the theory of heterogeneous nucleation at dislocations [25], as well as an RIS model [26, 27] used as a semi-empirical approach for tracking the dose (and dose rate)-dependent solute enrichment near dislocations.

*2.1 Heterogeneous nucleation on dislocations*

Dislocations are known to possess a catalytic effect on nucleation and act as favorable nucleation sites if the process releases their excess free energy. We assume that the nucleation of MNSPs follows the incoherent nucleation theory first developed by Cahn [25] and extended by Gomez-Ramirez and Pound [28]. The theory allows the nucleation of clusters to replace the region of dislocation cores while creating an incoherent interface with the matrix. The replaced dislocation segment is assumed to be smeared out as a continuous distribution of infinitesimal interfacial dislocations so as to reduce any strain energy contribution to near zero [28]. Thus the energy released by nucleation includes the contributions of dislocation core and strain energy inside the precipitate region. By applying this model and assuming the dislocation line passes through the center of the spherical MNSPs, the released excess free energy can be calculated. Additionally, dislocations provide a fast pipe diffusion pathway to the precipitates for solutes segregated to the dislocation lines. Thus we consider the rate controlling process for solute transport to be both 3D diffusion to the precipitates and 2D diffusion to the attached dislocation segments. Detailed calculations and parameters used in the model are described in SI Section S1.1 and S2.

We make the approximation that the number density of available heterogeneous nucleation sites at dislocations is equal to the number density of atoms along the dislocation line, and we assume that no new clusters nucleate within a distance of 5 nm from any evolving cluster to avoid precipitate volume overlap. This distance of 5 nm is chosen to be close to as small as possible given that the typical precipitate radius is ≈ 2 nm. Note that the separation distance, along with the dislocation density, sets the ultimate number density of precipitates. Thus for any value near 5 nm the number density predictions, for a specified dislocation density, will be relatively similar, and the exact predicted value is not particularly significant. Note we do not consider precipitate coherency strains, since the radiation-induced MNSP (G-phase) precipitates have a cube-on-cube orientation relationship with ferrite matrix with small lattice mismatch [29]. Any such strains would likely be compensated for by reductions in strains for a dislocation outside the precipitate region. In the present study we did not include heterogeneous nucleation at grain boundaries because of the lower nucleation site densities compared to dislocations, as detailed in SI Section S2. The recent APT studies of T91 strongly suggest that dislocations are the major nucleation site [4, 11]. More rapid nucleation is also expected due to the stronger RIS at dislocations than at grain boundaries in T91 [3, 4].

*2.2 Radiation-enhanced diffusion*

The RED model developed by Odette *et al*. [24] was used to calculate $X_vD_v$ and scale thermal diffusion coefficients. Under irradiation $X_v$ is much greater than the equilibrium vacancy concentration, $X_{ve}$, resulting in RED. Odette's model treats the effect of dpa-rate-dependent solute vacancy trapping that enhances recombination with self-interstitial atoms (SIA), reducing $X_vD_v$, relative to the condition when all the diffusing vacancies and SIA annihilate at sinks (no recombination) [24]. The effect of solute trap enhanced recombination is to increase the dpa required to reach a specified amount of precipitation under RED. In the RED calculation, we consider the highly concentrated Cr atoms in T91 as a solute traps that enhance with a binding energy of 0.094 eV taken from *ab-initio* calculation in Ref. [30]. Detailed $X_vD_v$ calculations are presented in SI Section S1.2.

*2.3 Radiation-induced segregation*

Heterogeneous nucleation at dislocations was treated by combining the local microalloy time (dose)-dependent RIS and cluster dynamic models. The RIS model is based on the formulations proposed by Wiedersich *et al*. [26] and Wolfer [27], that consider both the contributions of the inverse Kirkendall effect and vacancy drag as detailed in SI Section S3. The underlying mechanism driving Si RIS in our model is vacancy solute drag based on recent *ab initio* evaluation of transport coefficients in dilute bcc Fe-Ni and Fe-Si alloys [31]. Note, in contrast to Ni and Si, Mn segregation to dislocations is not observed in T91 [3]. Thus here only Ni and Si segregation was modeled. In both RIS calculations we did not include the contribution of grain boundaries as sinks due to the relatively much higher value of dislocation density. The grain boundary sink strength [32] of the 1-µm grain ($\sim 10^{13}$ m$^{-2}$) is about 2 orders of magnitude smaller than the dislocation sink strength.

*2.4 Description of simulations*

We first employ the thermodynamic simulations showing the equilibrium phase fraction of MNSPs in T91 at 400 °C, followed by the cluster dynamics simulations of Mn-Ni-Si precipitate formation induced by proton irradiation at 400 °C and $10^{-5}$ dpa/s. Figure 1 shows the flow chart of the cluster dynamics simulations, which involve the quantitative calculations of radiation and dislocation effects. The cluster dynamics model is calculated by a series of master equations of cluster distribution functions (Eq. (S1)) in which their evolution is determined by the rate coefficients and formation energy of clusters. The former includes the contributions of thermal diffusion, radiation-induced diffusion by the diffusion paths through the matrix and dislocation lines, which are detailed in SI Section S. The latter is calculated by Eq. (S15) which considers the chemical free energy, interfacial energy, and catalytic effect of dislocations. It is noted that for the simulation of heterogeneous nucleation at dislocations, the free energy contribution is calculated by using the RIS composition at the segregated and microalloyed region near dislocations.

The CD model integrates the RIS and RED calculation results of proton-irradiated alloy T91 at 400 °C and a dose rate of $10^{-5}$ dpa/s. The model considers multiple dislocation effects, including the density of heterogeneous nucleation sites and the decrease of nucleation free energy barrier. The nucleation barrier was calculated based on the free energy reduction driving phase separation in the RIS microalloy, the precipitate-matrix interfacial energy and the energy release associated with annihilation of the dislocation segment inside the precipitate. The dpa-dependent local RIS compositions (enrichment) of Ni and Si scaled in proportion to the instantaneous solute concentrations in matrix. The bulk solute contents decreased with dpa in proportion to the Mn, Ni, Si precipitation. The cluster dynamics model was used to simulate the evolution of the number density ($N$), mean radius ($<r>$) and mole fraction ($f$) of MNSPs at 400°C up to 100 dpa.

## 3 Results and discussion

*3.1 Thermodynamic calculations*

The thermodynamic state in T91 was evaluated first by the Thermo-Calc software using the TCAL3 database [33]. Three alloy compositions were considered: the original T91 (a); the RIS composition measured by Wharry *et al*. [4] at grain boundaries (b); and, the composition measured by Jiao *et al.* [3] at dislocations (c). The measured compositions were 0.45 at% Mn, 0.58 at% Ni and 0.95 at% Si at grain boundaries, and 0.37 at% Mn, 0.78 at% Ni, and 4.01 at% Si at dislocations. Figure 2 shows the calculated phase fraction of the equilibrium bulk MNSPs (no interface effects) as a function of temperature. Only the T3 ($Mn_6Ni_{16}Si_7$) phase (also called G-phase) is stable. In the case of the bulk composition prior to irradiation, MNSPs are not able to form at temperatures higher than 327 °C. However, for the RIS compositions, the maximum temperature for the formation of MNSPs increases to 410 °C at grain boundaries and 505 °C at dislocations. The corresponding thermodynamic phase fraction is 0.12% and 1.0% at 400 °C at grain boundaries and dislocations, respectively. Note that the phase fraction of precipitate predicted here would only occur in the RIS enhanced region, which can be considered a local "microalloy", and not throughout the bulk of the alloy [34]. These results suggest that the formation of the MNSPs is radiation-induced by the segregation of solutes at defect sinks, which increases the local driving force of precipitate nucleation and growth. The thermodynamic prediction is consistent with the experimental observation [4] showing MNSPs in proton-irradiated T91 at 400 °C. Note that in this thermodynamic analysis, the T6 ($Mn(Ni,Si)_2$) phase was included in the calculation but did not form even for the T91 local dislocation RIS composition. Therefore, the precipitation modeling of MNSPs in this study focuses on the T3 or G-phase, which are also observed in RPV steels [6-9] and HT-9 F-M alloy [35, 36].

*3.2 Result of radiation-induced segregation*

Figure 3 shows the predicted RIS as a function of dpa dose at 400 °C for the experimental condition of ≈ $10^{-5}$ dpa/s for the proton irradiations in Ref. [3]. The solid lines are simple fits to the small filled square computed data points, while the large open symbols are the experimentally observed solute concentration at dislocations. RIS increases up to a steady state local concentration of Ni ≈ $7 \times 10^{-3}$ at.% at about 4 dpa, while the Si segregation does not saturate at less than 10 dpa, where it reaches the observed concentration of ≈ 4.5%. The fitting to experimental data [4] was done by adjusting the pre-exponential factors to find the consistent magnitude of RIS integrated through a distance of 2 nm from the dislocation core. The RIS model is used to estimate the local microalloy composition at dislocations, hence the chemical driving force of precipitation, as a function of dpa. We note that the RIS model is very approximate and cannot be used for quantitative prediction outside the conditions studied here (please see SI Section 3 for more details).

*3.3 Cluster dynamic simulation results*

The result of RIS and RED calculations are integrated in the CD model for the simulation of MNSPs in proton-irradiated alloy T91 at 400 °C and a dose rate of $10^{-5}$ dpa/s, including the evolution of the number density (*N*), mean radius (<*r*>) and mole fraction (*f*) of MNSPs up to 100 dpa.

The predicted *N*, <*r*> and *f*, respectively are shown in Figure 4 (a), (b) and (c). Figure 4 also shows the corresponding experimental data reported by Jiao *et al.* [11]. Note that to enable a direct comparison with the G-phase MNSPs in the model, the radii from experiments were estimated based only on the Mn, Ni, and Si atomic fractions and volumes in a precipitate, and any nominal contributions from Fe were removed. This approach is consistent with evidence that the Fe in the MNSPs is an artifact of the APT [11]. The volume is estimated from the crystal structure and number of atoms in a unit cell of G-phase. Only the clusters with sizes larger than 65 atoms were counted in the calculations of number density and mean radius, which is consistent with the typical resolution limit of the APT. The result in Figure 4 demonstrates excellent consistency with the experimental observation [11] in number density and mean radius, which were characterized as $1.27\times10^{23}$ m$^{-3}$ and 1.6 nm, respectively. The model predicts that rapid nucleation starts at doses between 0.2 and 0.5 dpa, and the volume fraction grows continuously after about 1 dpa. This evolution of the MNSPs is also in qualitative agreement with the description by Jiao *et al.* [11] who reported the observable MNSPs started to appear at doses between 1 and 7 dpa. However it is again important to note that the choice of the minimum precipitate spacing and the adjusting of parameters in the RED model (see S.I. Sec. 1.2) ensure that the predictions are consistent with the proton data if the underlying thermodynamic assumptions and segregation estimates are valid.

To develop a clearer understanding of dislocation and radiation effects on G-phase precipitation, we explore the separate effects of dislocations, RIS, and RED on the evolution of precipitation based the same framework of model and parameters. Figure 5 (a) and (b) shows respectively the simulation results of $N$, $<r>$ and $f$ for various combinations of mechanisms, including the full calculations (with dislocations, RIS and RED) and without RIS or RED. The results show that as long as dislocations are included as a favorable nucleation site, then $N \approx 3.5 \times 10^{21}$ m$^{-3}$ at 7 dpa. The absence of RIS lowers $N$ relative to the full model, but the more significant effect of the absence of RIS is that the precipitates do not grow larger than 0.59 nm in mean radius (an average of just 70 atoms), which is much smaller than the reported mean radius (1.6 nm). Indeed in this case the number of atoms in the precipitates is less than a single G-phase unit cell (116 atoms) with a lattice parameter of 1.117 nm [37], hence, are better described as slowly evolving clusters rather than precipitates. These small clusters are stabilized by the nominal energy gain associated with annihilation of the dislocation core, do not grow in absence of the RIS. Given that the model makes other significant approximations with respect to very small clusters on the dislocations, e.g., ignoring any thermal segregation and assuming complete annihilation of the dislocation core, the model prediction might deviate significantly from actual cluster behavior in this situation. However, regardless of the detailed accuracy of the model in this somewhat artificial limit of no RIS, the results indicate that segregation is necessary to provide sufficient driving force for MNSP nucleation and growth. In contrast to RIS, RED simply accelerates the precipitate evolution, shifting it to a lower dpa.

While the present model has been successful in its goal of determining the qualitative mechanisms controlling precipitation of MNSPs in T91 under proton irradiation, it cannot yet provide quantitative predictions for other alloy and irradiation conditions. Refinements that might lead to more generally quantitative predictions, including more accurate modeling of RIS and other relevant physics like dislocation loop evolution. Additionally, Wharry [4] have reported that about 76% of the MNSPs are associated with CRPs in T91. These authors hypothesize that CRPs formed after the MNSPs since some of the MNSPs were not associated with CRPs. However, studies of neutron-irradiated Cu-Ni-Mn-Si RPV steels show that CRPs form much earlier than MNSPs. The CRPs enhance the nucleation and growth of MNSPs by providing sites to form co-precipitate appendages [7, 38, 39]. Notably at low supersaturations CRPs also form preferentially on dislocations, as illustrated in the APT reconstruction in SI Section S5. The synergistic effects of dislocations and CRPs on MNSP evolution should be the focus of future work.

## 4 Conclusions

In summary, we have developed a model to study the evolution of MNSPs in F-M alloy T91 under proton irradiation. The approach is based on a CD model with heterogeneous nucleation on dislocations, including the effects of RIS. Dose-dependent local microalloy solute concentrations are estimated based on a fitted continuum RIS model. The local solute concentrations are used in the CD calculation. In the absence of dislocation effects on heterogeneous nucleation and RIS, the CD model underestimates the number density and size of the MNSPs compared to the proton irradiation results. In contrast, including the catalytic effect of dislocations and RIS, with all fitted parameters in physically reasonable ranges, the CD model is consistent with observations, although this is partially imposed, like the saturated *N*. The model developed here can be used to qualitatively explore the effects of temperature and dose rate on MNSPs precipitation. As detailed in SI Section S6, decreasing the dpa rate by a factor of 100 at 400 °C, closer to in-service neutron irradiation conditions, decreases the MNSP volume fraction by ~35%. Decreasing the temperature to 300 °C at a dpa rate of $10^{-5}$ dpa/s increases the volume fraction by ~43%. However, a number of improvements in the model are needed for more quantitatively reliable predictions, including more physics as discussed above and better constrained fitting parameters as described in SI Section S4. The physically reasonable values obtained for all the fitting parameters suggests that the model is representative of the dominant physics in the problem, but the extensive fitting to limited data means that the model cannot be used for more than qualitative guidance.

The model supports the hypothesis that MNSPs in T91 are controlled by the combination of G-phase thermodynamics, RIS and dislocation enhanced heterogeneous nucleation rates, consistent with previous observations and interpretations [4, 11]. The approach in this study provides a framework for integrating RIS and dislocation effects into more general modeling of precipitation in under-saturated alloys with realistic treatment of their microstructures.

**Acknowledgements**

This work at the University of Wisconsin-Madison and UC Santa Barbara (UCSB) is funded by the US Department of Energy (DOE), Office of Nuclear Energy (NE), Integrated Research Project (IRP) titled "High Fidelity Ion Beam Simulation of High Dose Neutron Irradiation" (DE-NE0000639). Partial support at UCSB was provided by the DOE Office of Fusion Energy Sciences Grant DE-FG03-94ER54275.

**FIGURE CAPTION**

Figure 1. Flow chart showing the cluster dynamics simulation of MNSPs formation in T91 under irradiation.

Figure 2. Phase fraction of the T3 or G-phase as a function of temperature calculated by TCAL3 database. The black, blue and orange curves show the calculated results of T91 original and RIS compositions at grain boundaries (GB) and dislocations (DL), respectively. No other MNSPs are predicted to be stable.

Figure 3. Calculation result showing the evolution of Ni and Si RIS compositions at dislocations as a function of dose. The symbols show the experimental values reported in Ref. [3].

Figure 4. Calculation result of the cluster dynamics model showing the (a) number density, (b) mean radius, and (c) volume fraction of MNSPs as a function of irradiation dose (dpa). The dose rate and temperature are $10^{-5}$ dpa/s and 400 °C, respectively. The model includes the effect of heterogeneous nucleation at dislocations as well as the evolution of RIS shown in Figure 2. The symbols show the values reported by the experiment [11]. The error bar on the experimental radius indicates the standard error in the mean radius ($\sigma_{\langle r \rangle}$) determined by the formula $\sigma_{\langle r \rangle} = \sigma_r/\sqrt{N}$, where $\sigma_r$ is the standard deviation of the precipitate size of all the APT-identified particles and $N = 43$ is the number of particles that were measured.

Figure 5. Calculation result of the cluster dynamics model showing the (a) number density and (b) mean radius of MNSPs as a function of irradiation dose (dpa) under various conditions, including the full calculations (with dislocations, RIS and RED) and calculations without RIS or RED. The dose rate and temperature are $10^{-5}$ dpa/s and 400 °C, respectively. The symbols show the values reported by the experiment [11].

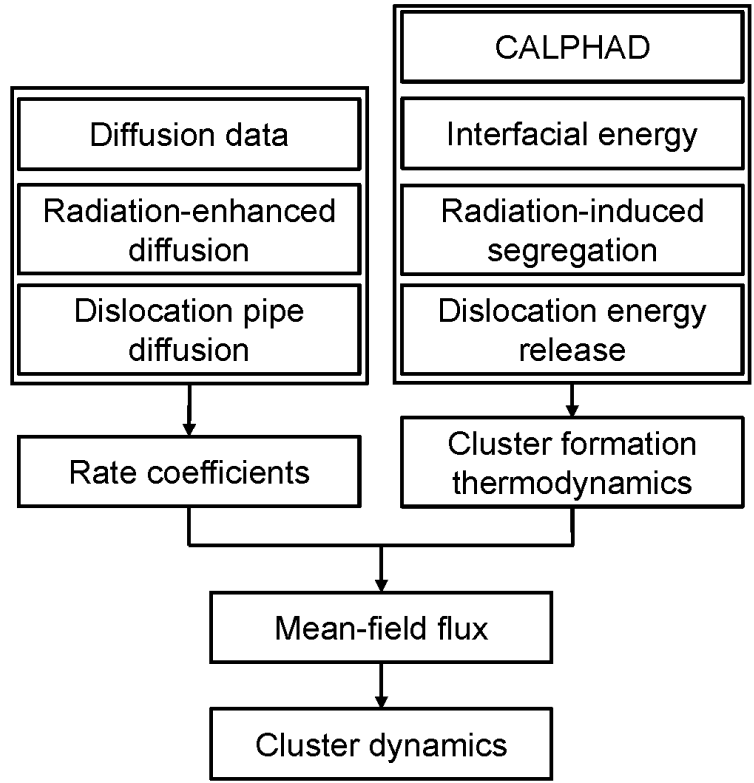

Figure 1. Flow chart showing the cluster dynamics simulation of MNSPs formation in T91 under irradiation.

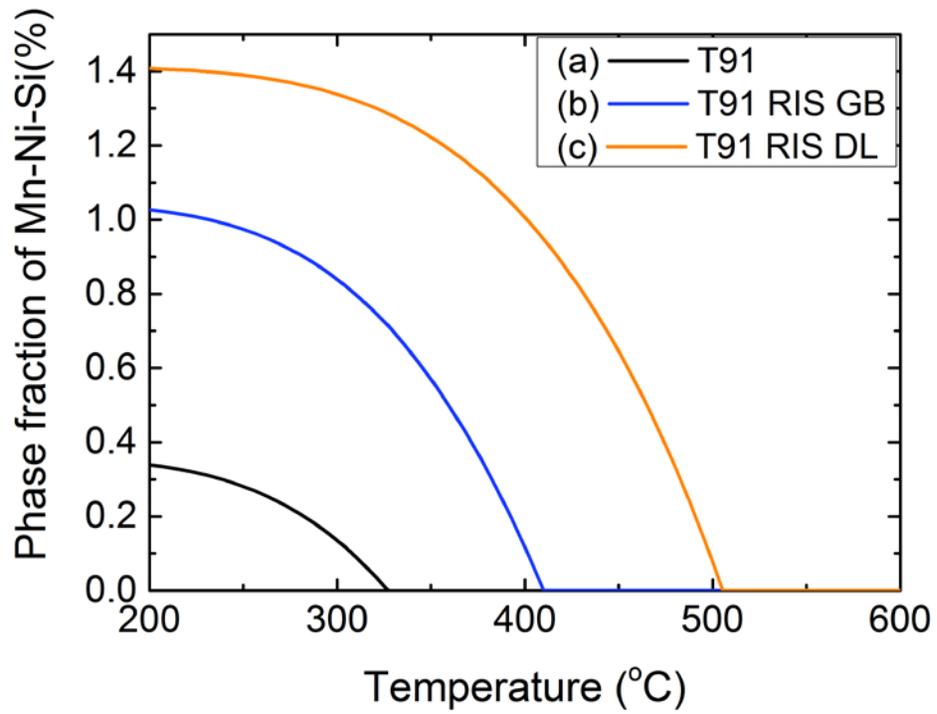

Figure 2. Phase fraction of the T3 or G-phase as a function of temperature calculated by TCAL3 database. The black, blue and orange curves show the calculated results of T91 original and RIS compositions at grain boundaries (GB) and dislocations (DL), respectively. No other MNSPs are predicted to be stable.

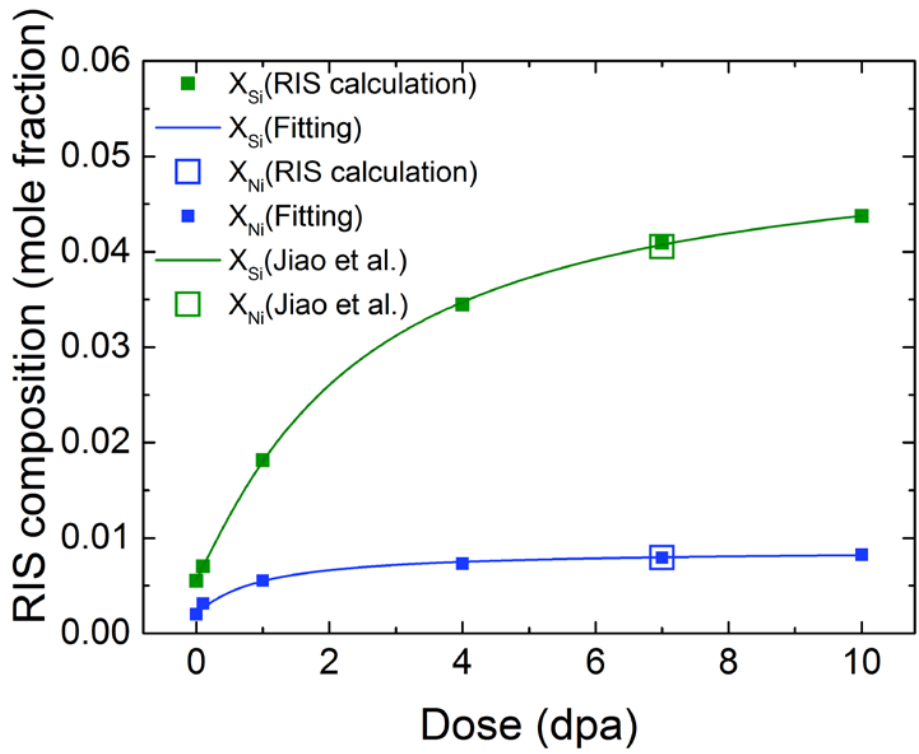

Figure 3. Calculation result showing the evolution of Ni and Si RIS compositions at dislocations as a function of dose. The symbols show the experimental values reported in Ref. [3].

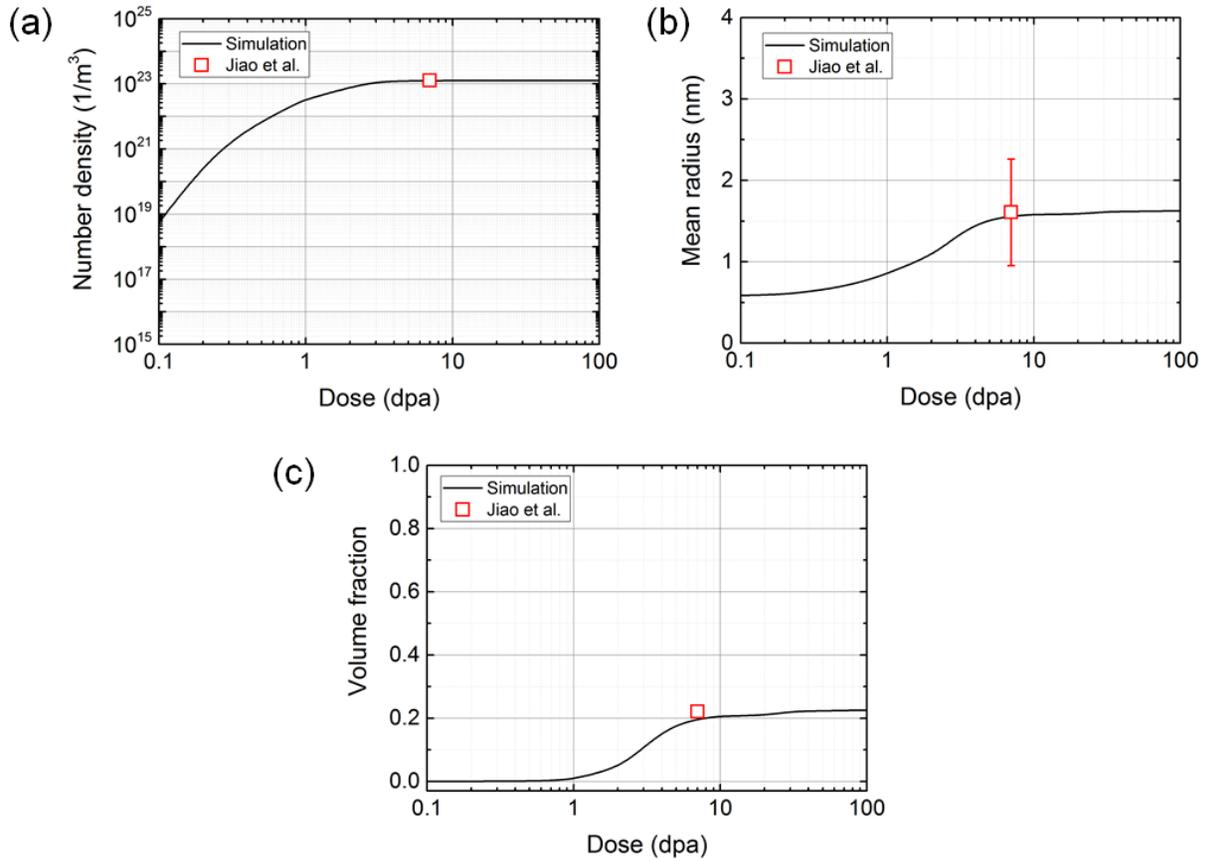

Figure 4. Calculation result of the cluster dynamics model showing the (a) number density, (b) mean radius, and (c) volume fraction of MNSPs as a function of irradiation dose (dpa). The dose rate and temperature are $10^{-5}$ dpa/s and 400 °C, respectively. The model includes the effect of heterogeneous nucleation at dislocations as well as the evolution of RIS shown in Figure 2. The symbols show the values reported by the experiment [11]. The error bar on the experimental radius indicates the standard error in the mean radius ($\sigma_{\langle r \rangle}$) determined by the formula $\sigma_{\langle r \rangle} = \sigma_r/\sqrt{N}$, where $\sigma_r$ is the standard deviation of the precipitate size of all the APT-identified particles and $N = 43$ is the number of particles that were measured.

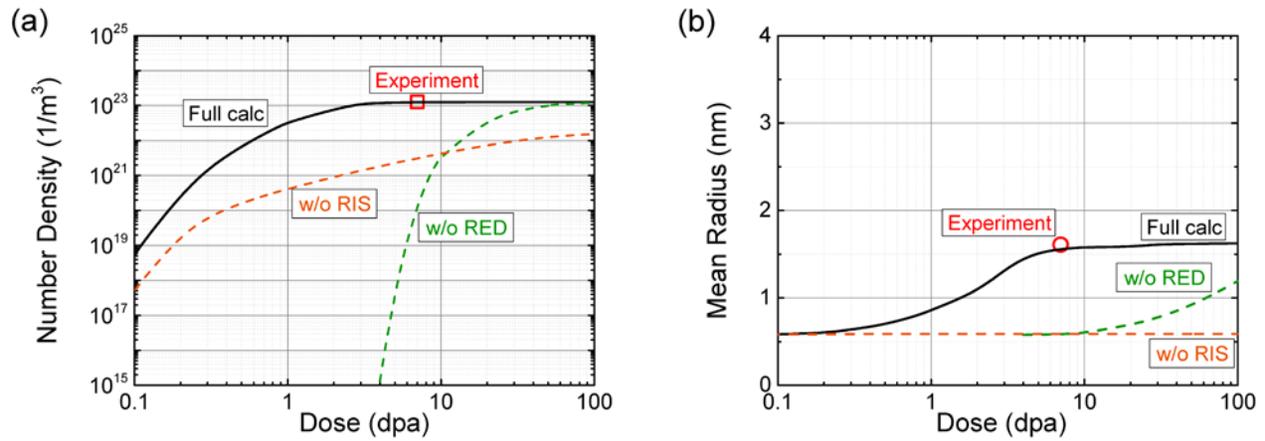

Figure 5. Calculation result of the cluster dynamics model showing the (a) number density and (b) mean radius of MNSPs as a function of irradiation dose (dpa) under various conditions, including the full calculations (with dislocations, RIS and RED) and calculations without RIS or RED. The dose rate and temperature are $10^{-5}$ dpa/s and 400 °C, respectively. The symbols show the values reported by the experiment [11].

# Supplementary Information

**Cluster dynamics modeling of Mn-Ni-Si precipitates in ferritic-martensitic steel under irradiation**

Jia-Hong Ke [a], Huibin Ke [a], G. Robert Odette [b] and Dane Morgan [a*]

[a] Department of Materials Science and Engineering, University of Wisconsin-Madison, Madison, WI 53706, USA

[b] Materials Department, University of California, Santa Barbara, CA 93106, USA

* Corresponding author. E-mail address: ddmorgan@wisc.edu

## S1. Cluster dynamics modeling of Mn-Ni-Si-rich precipitates

*S1.1 Cluster dynamics model*

We followed the kinetic equations developed by Slezov and Schmelzer [1-4] to describe nucleation-growth processes of diffusional phase transformations in multicomponent systems. The model assumes that the Mn-Ni-Si-rich phases can be treated as pure Mn-Ni-Si phases, with no other alloying elements. With the assumption that only monomers can migrate, the discrete cluster size distributions are governed by the coupled master equations:

$$\frac{\partial f(n,t)}{\partial t} = J_{n-1} - J_n \tag{S0}$$

where $f$ is the distribution function of clusters with $n$ atoms at time $t$ and $J_n$ is the flux of clusters between cluster sizes $n$ and $n + 1$. The flux can be determined by

$$J_n = \omega_{n,n+1}^{(+)} f(n,t) - \omega_{n+1,n}^{(-)} f(n+1,t) \tag{S1}$$

where $\omega_{n,n+1}^{(+)}$ is the absorption coefficient or the rate at which a cluster of size $n$ absorbs a single atom and grow to size $n$ +1, and similarly, $\omega_{n+1,n}^{(-)}$ is the emission coefficient or the rate at which a cluster of size $n$ + 1 emits a single particle and shrinks to size $n$. The symbols (+) and (−) correspond to absorption and emission of an atom or monomer, respectively. It is noted that in the present study, we adopt the treatment from Slezov and Schmelzer [4] by assuming that the evolution of clusters follows the trajectory of minimum free energy in composition space, and accordingly the cluster size distribution considered here is a function of the total number of atoms in the cluster (instead of the number of each components in the cluster). The absorption coefficient can be determined directly by the macroscopic growth kinetics of the cluster. With the assumptions that the growth of clusters is diffusion-limited and spherical in shape, the effective absorption coefficient can be expressed as [4]

$$\omega_{n,n+1}^{(+)} = 4\pi c_\beta a_\alpha D_{\text{eff}}^d n^{1/3} \tag{S2}$$

and

$$\frac{1}{D_{\text{eff}}^d} = \sum_i \frac{v_{i\alpha}^2}{x_{i\beta} D_i} \tag{S3}$$

where $c_\beta$ and $x_{i\beta}$ are the total volume concentration of the particles and molar fraction of the component $i$ in the ambient phase $\beta$, respectively, and $D_i$ is the diffusion coefficient of species $i$ in the ambient phase. A fixed atomic fraction in the clusters of precipitates is assumed, and the parameter $v_{i\alpha}$ is the atomic fraction of the component $i$ in the precipitate phase $\alpha$.

The absorption coefficient in Eq. (S3) is derived under the assumption that the growth is controlled by volume diffusion and the long-range matrix composition is located far away from the cluster interphase boundaries. The assumption will not be valid for a microstructure affected by other diffusion pathways or local solute segregation. Modification of Eq. (S3) is necessary for the precipitation of Mn-Ni-Si-rich phase in T91 under proton irradiation, which was found to be strongly associated with dislocations and radiation-induced segregation [5]. Particularly, because of the extremely fast diffusion along dislocation pipes, dislocations provide an additional diffusion pathway facilitating the nucleation and growth of precipitates. We assume that the diffusion flux along dislocations is controlled by the long-range diffusion into the dislocation lines, and that any solutes arriving at the dislocation are immediately added to the precipitate. Therefore, the absorption coefficient should also consider this additional dislocation transport contribution, which is cylindrical in symmetry owing to the diffusion toward the straight dislocation lines. By considering pathways of volume diffusion and transport along dislocation pipes, the modified absorption coefficient is derived as

$$\omega_{n,n+1}^{(+)} = \begin{cases} 2\pi c_\beta D_{\text{eff}}^d \dfrac{d - 2r_p^n}{\ln(r_0/r_{\text{core}})} & \text{for } r_p^n < r_{\text{core}} \\ 4\pi c_\beta D_{\text{eff}}^d \dfrac{r_p(n) r_0}{r_0 - r_p^n} \dfrac{\sqrt{(r_p^n)^2 - r_{\text{core}}^2}}{r_p^n} + 2\pi c_\beta D_{\text{eff}}^d \dfrac{d - 2r_p^n}{\ln(r_0/r_{\text{core}})} & \text{for } r_p^n > r_{\text{core}} \end{cases} \quad (S3)$$

where $r_p^n$ is the radius of a cluster with $n$ atoms, $r_0$ is the radius of the microalloy region with solute segregation, $r_{\text{core}}$ (0.4 nm [6]) is the dislocation core radius, and $d$ is the distance which avoids the overlap of precipitate volume on dislocations, which is 5 nm in this study. A recent study of radiation-induced segregation (RIS) in T91 shows that the RIS-affected microalloyed region is approximated 5 nm in size [7], so we take $r_0 = 2.5$ nm in Eq. (S5). For precipitates with a radius smaller than the dislocation core, diffusion along dislocation pipes dominates (Eq. (S5, top)), whereas for that precipitates larger than the dislocation core, both volume diffusion and transport along dislocation pipes ((Eq. (S5, bottom)) contribute to the growth of precipitates. The first term in Eq. (S5, bottom) reflects the cluster growth by volume diffusion and the ratio $\sqrt{r_p^2 - r_c^2}/r_p$ removes the overlapped contribution of the surface area intersecting with dislocation cores. The second term corresponds to the cylindrical diffusion into the dislocation lines, which takes effect for precipitates on dislocations with any given size.

The emission coefficient can be expressed by the given relation:

$$\omega_{n+1,n}^{(-)} = \omega_{n,n+1}^{(+)} \exp\left(\dfrac{\Delta G(n+1) - \Delta G(n)}{kT}\right) \quad (S4)$$

where $\Delta G(n)$ is the formation free energy of a cluster containing n atoms. Note that, consistent with the assumption of very fast dislocation transport used above, the dislocation is in equilibrium with the surrounding bulk and we need only consider changes in Gibbs free energy associated with emission into the bulk region near the dislocations in Eq. (S6). If only the chemical free energy and interfacial energy are considered, $\Delta G(n)$ can be expressed as

$$\Delta G(n) = n\left(g_p - \sum_i x_i \mu_i\right) + 4\pi\sigma \left(\frac{3n\Omega_p}{4\pi}\right)^{2/3} \qquad (S5)$$

$g_p$ is the free energy per atom of the precipitate phase, $\mu_i$ is the chemical potential of component $i$ in the matrix, $\sigma$ is the interfacial energy of the precipitate per unit area, and $\Omega_p$ is the atomic volume of the precipitate phase. For dilute components such as the Mn, Ni, and Si additions in the alloy T91 as listed in Table S1, $g_p - \sum_i x_i \mu_i$ can be simplified to the expression of solute products $\overline{\prod_i c_i^{x_i}} / \prod_i c_i^{x_i}$, in which the bar on the solute product indicates the value in thermodynamic equilibrium and can be obtained from thermodynamic database. In our thermodynamic model based on the TCAL3 database the only Mn-Ni-Si phase we find stable for T91 (at normal bulk or RIS compositions) is the T3 (or G-phase), with a composition of $Mn_6Ni_{16}Si_7$ [8]. We therefore focus on this phase in the cluster dynamics modeling. The value of $\overline{\prod_i c_i^{x_i}}$ for the Mn-Ni-Si phase in T91 at 400 °C can be obtained as $6.26\times10^{-3}$ according to the TCAL3 database, where $x_i$ for Mn, Ni, and Si are respectively 0.21, 0.55, and 0.24. The interfacial energy was assumed to be 0.19 J m$^{-2}$ according to the fitting result for the G-phase in RPV steels [9].

Table S1. Chemical composition of T91 in at% and wt% [10].

|       | Cr   | Ni   | Mn   | Si   | C    | P     | Cu   | V     | Mo   | S     | N     | Nb    | Al    |
|-------|------|------|------|------|------|-------|------|-------|------|-------|-------|-------|-------|
| at.%  | 8.90 | 0.20 | 0.45 | 0.55 | 0.46 | 0.016 | 0.15 | 0.23  | 0.52 | 0.005 | 0.19  | 0.005 | 0.045 |
| wt.%  | 8.37 | 0.21 | 0.45 | 0.28 | 0.1  | 0.009 | 0.17 | 0.216 | 0.9  | 0.003 | 0.048 | 0.008 | 0.022 |

*S1.2 Radiation-enhanced diffusion*

We apply the radiation-enhanced diffusion model developed by Odette *et al*. [11] to calculate $X_vD_v$ and scale thermal diffusion coefficients. The radiation enhanced diffusion coefficients is given as

$$D_i^{rad} = X_V^r \frac{D_i^{th}}{X_V^e} + D_i^{th} \qquad (S6)$$

where $X_V^r$ is the non-equilibrium vacancy concentration under irradiation, $X_V^e$ is the vacancy concentration at thermodynamic equilibrium. $D_i^{th}$ is the diffusion coefficient of the solute $i$ under the condition of thermal annealing. Under the steady state when defect production is balanced by annihilation at sinks as well as recombination at matrix and solute trapped vacancies, the vacancy concentration can be expressed as

$$X_V^r = \frac{g_s \xi \sigma_{dpa} \phi}{D_V S_t} \qquad (S7)$$

where $\sigma_{dpa}$ is the displacement-per-atom (dpa) cross-section, $\xi$ is the cascade efficiency or the fraction of vacancies and self-interstitial atom (SIA) created per dpa, $S_t$ is the total sink strength, and $r_v$ is the SIA-vacancy recombination radius. The total sink strength includes the contribution of dislocations ($S_d$) and vacancy clusters ($S_c$). The former can be characterized by the dislocation density and the latter can be evaluated by

$$S_c = 4\pi r_c \sigma_c \phi \tau_c / \Omega_a \qquad (S7)$$

where $r_c$, $\sigma_c$ and $\tau_c$ are the recombination radius, production cross-section and annealing time for vacancy clusters. $g_s$ is the vacancy survival fraction which can be obtained by solving the steady-state equation [11]:

$$1 - \frac{R_r \xi \sigma_{dpa} \phi (g_s)^2}{D_i D_v S_t^2} - g_s - \frac{\xi \sigma_{dpa} \phi R_t^2 X_t \tau_t (g_s)^2}{S_t^2 \left(1 + g_s \xi \sigma_{dpa} \phi R_t \tau_t / S_t\right)} = 0 \qquad (S7)$$

$R_r$ and $R_t$ are respectively the matrix and trap recombination radii, $X_t$ is the trap density, and $\tau_t$ is the annealing time for trapped vacancies which can be expressed as [11]

$$\tau_t = d^2 \Big/ D_V e^{-\frac{H_b}{kT}} \qquad (S8)$$

where $H_b$ is the binding energy for trapped vacancies and $d$ is the nearest neighbor distance of bcc Fe lattice ($2.48 \times 10^{-10}$ m). Here we consider the highly concentrated Cr atoms in T91 (9%) as the main trapping site. Table S2 lists the parameters used in the calculation of radiation-enhanced diffusion.

Table S2. Parameters used in the RED model for calculating radiation-enhanced diffusion. * The variables marked with a *, which include "Vacancy cluster recombination radius" and "Vacancy cluster annealing time" were fit to match the experimental onset of MNSP after 1 dpa and the observed size and number density at 7 dpa, as reported in Ref. [5].

| Parameter | Symbol | Value |
|---|---|---|
| SIA – vacancy recombination radius | $r_v$ | 0.57 nm [9, 11] |
| *Vacancy cluster recombination radius | $r_c$ | $5.0 \times 10^{-10}$ m |
| Displacement-per-atom cross-section | $\sigma_{dpa}$ | $1.5 \times 10^{-25}$ m$^2$ [9, 11] |
| Vacancy cluster production cross-section | $\sigma_c$ | $4.5 \times 10^{-25}$ m$^2$ |
| *Vacancy cluster annealing time | $\tau_c$ | $8.1 \times 10^{-12}/e^{(-1.85/kT)}$ |
| Cascade efficiency | $\xi$ | 0.9 [12] |
| Atomic volume | $\Omega_a$ | $1.18 \times 10^{-29}$ m$^3$ [9] |
| Vacancy diffusion coefficient at 400 °C | $D_V$ | $5.79 \times 10^{-13}$ m$^2$ s$^{-1}$ [9] |
| Dislocation sink strength (dislocation density) | $S_d$ | $6.25 \times 10^{14}$ m$^{-2}$ [13] |
| Fe self-diffusivity at 400 °C | $D_{Fe}^{th}$ | $5.33 \times 10^{-23}$ m$^2$ s$^{-1}$ [9] |
| Mn diffusivity in Fe at 400 °C | $D_{Mn}^{th}$ | $1.02 \times 10^{-22}$ m$^2$ s$^{-1}$ [9] |
| Ni diffusivity in Fe at 400 °C | $D_{Ni}^{th}$ | $1.18 \times 10^{-23}$ m$^2$ s$^{-1}$ [9] |
| Si diffusivity in Fe at 400 °C | $D_{Si}^{th}$ | $8.36 \times 10^{-23}$ m$^2$ s$^{-1}$ [9] |
| Matrix recombination radius | $R_r$ | $5.7 \times 10^{-10}$ m [11] |
| Trap recombination radius | $R_t$ | $5.7 \times 10^{-10}$ m [11] |
| Trap concentration | $X_t$ | 0.09 |
| Binding energy for trapped vacancies | $H_b$ | 0.094 eV [14] |

## S2. Consideration of heterogeneous nucleation at dislocations in cluster dynamics model

Following the model by Cahn [15] and Gomez-Ramirez [16] and assuming the dislocation line

passes through the center of the spherical Mn-Ni-Si precipitate or cluster, the released excess free energy associated with the nucleation of a cluster at a dislocation can be given as:[15]

$$\Delta G_{disl}\left(r_p\right) = \begin{cases} \int_{-r_p}^{r_p} [E_{core}] \, dl & \text{precipitate on core, } r_p < r_{core} \\ \int_{-r_p}^{r_p} [E_{core} + \dfrac{Aa_0^2}{4\pi} \ln(r(l)/r_{core})] \, dl & \text{precipitate on core, } r_p > r_{core} \end{cases} \quad \text{(S8)}$$

$r_{core}$ and $E_{core}$ are respectively the dislocation core radius and core energy, whose values we take from the estimations by Marian et al. [17] and Dudarev et al. [6]. $r_p$ is the cluster radius, $r$ is the distance between a point on the precipitate interface and the dislocation line, $l$ is the distance from the center of a precipitate along the dislocation line, $a_0$ is the lattice constant ($2.87 \times 10^{-10}$ m), and $A$ is a factor that corresponds to the anisotropic elastic strain energy of dislocations [18]:

$$A = K b^2 / a_0^2 \quad \text{(S8)}$$

where $b$ is the dislocation Burgers vector and $K$ is the energy factor which depends on the elastic constant $c_{ij}$, dislocation type and direction ($l$). The factor $A$ can be calculated by following Ref.[18] for various types of dislocations and typical values are summarized in Table S3. In this study we consider the [001] type pure edge dislocation for all of our CD calculations with a value $A = 104.2$ GPa. It is noted that the variation of $A$ from 57.7 to 134.2 GPa does not have significant influence on the growth of MNSPs in both number density and size. The calculation results using the different $A$ factors (57.7, 134.2, and 104.2 GPa) are shown in Section S7, and indicate that larger $A$ causes only slightly earlier nucleation due to the larger energy release by dislocation strain energy. The insensitivity of the precipitate evolution to the $A$ value is reasonable in this study because as RIS develops and increases with dose and time, the nucleation barrier becomes much lower and the energy release by the dislocation strain energy produces relatively minor impact.

It should be noted that the two lines in Eq. (S13) are, from top to bottom, for clusters that are on the dislocation and have sizes smaller than the dislocation core and for clusters that are on the dislocation and that have size larger than the core radius, respectively.

Table S3. The factor $A$ (Eq. (14)) for various types of dislocation in bcc Fe.

| Dislocation type | $l$ (direction) | $b/a_0$ | $A$ (GPa) |
|---|---|---|---|
| Screw | [001] | [001] | 116.0 |
| Edge in {100} | [001] | [100] | 104.2 |
| Mixed in {110} | [001] | 1/2[111] | 81.1 |
| Mixed in {100} | [101] | [100] | 91.2 |
| Edge in {110} | [101] | [010] | 134.2 |
| Mixed in{110} | [101] | 1/2 [111] | 57.7 |
| Edge in {112} | [101] | 1/2 [111] | 100.7 |

By considering the catalytic effect of dislocation on nucleation, the total formation free energy of a cluster in the cluster dynamics model then becomes: [16]

$$\Delta G(n) = n\left(g_p - \sum_i x_i \mu_i\right) + 4\pi\sigma\left(\frac{3n\Omega}{4\pi}\right)^{2/3} + \Delta G_{\text{disl}}(r_p) \tag{S8}$$

The last term corresponds to the released excess free energy associated with the nucleation of a cluster at a dislocation as given in Eq. (S13). The dislocation core radius and core energy were estimated by Marian *et al*. [17] and Dudarev *et al*. [6]. The structural information of dislocation cores together with the reported dislocation density allows us to calculate nucleation on dislocations semi-quantitatively. Table S4 lists the parameters used in the calculations. Note that these parameters are for nominally pure bcc Fe as we are not aware of values for T91.

We note that heterogeneous nucleation at grain boundaries can be modeled in a similar manner as at dislocations, but it was not considered in the present study because recent APT studies strongly suggest that dislocations are the major nucleation site [5, 19]. This is expected, since the grain boundaries provide many fewer sites for nucleation than the dislocations. More specifically, based on the measured dislocation density of $\rho_d = 6.25\times10^{14}$ m$^{-2}$ and an effective grain size of 1.49 μm [13], the heterogeneous nucleation site densities can be estimated as $3.3 \times 10^{24}$ m$^{-3}$ and $6.3 \times 10^{23}$ m$^{-3}$ for dislocations and grain boundaries, respectively. Here we have assumed cubic grains with side dimensions equal to the grain size and that nucleation sites are separated by 5 nm (same as above) on a square grid with cell dimensions of 5 nm. Thus the heterogeneous nucleation site density is about 5 times larger for dislocations than grain boundaries. More rapid nucleation is also expected due to the stronger RIS at dislocations than at grain boundaries in T91 [7, 19].

Table S4. Parameters for bcc Fe that are used in the cluster dynamics model of heterogeneous nucleation.

| | |
|---|---|
| $r_{core}$ | 0.4 nm [6] |
| $E_{core}$ | 0.937 eV/Å [6] |
| $b = a_0$ | 0.287 nm [20] |
| $\Omega_a$ | $1.18 \times 10^{-29}$ m$^{-3}$ [20] |
| $c_{11}$ | 231.5 GPa [21] |
| $c_{12}$ | 135.0 GPa [21] |
| $c_{44}$ | 116.0 GPa [21] |
| $\rho_d$ | $6.25 \times 10^{14}$ m$^{-2}$ [13] |

**S3. Continuum modeling of radiation-induced segregation**

We implement the theory of RIS developed by Wiedersich *et al.* [22] and Wolfer [23], which considers both the contributions of the inverse Kirkendall effect and vacancy drag. We adapted the codes that were applied successfully to study the Cr segregation behavior of the 9 wt% Cr F-M steel [24]. The time evolution of the solute and defect concentrations are respectively given by

$$\frac{dC_A}{dt} = -\frac{dJ_A}{dx} \tag{S8}$$

and

$$\frac{dC_d}{dt} = -\frac{dJ_d}{dx} + \xi K_0 - R C_I C_V \tag{S8}$$

where the subscript $A$ denotes the alloy component, $d$ is the type of defects such as vacancies and interstitials, $J$ is the diffusion flux, $K_0$ is the dpa rate of irradiation, $\xi$ is the displacement efficiency, and $C_I$ and $C_V$ are concentrations of interstitials and vacancies. By following the treatment of Ref. [23] and assuming a dilute system with negligible dependence of of thermodynamic factor and vacancy formation energy on alloy composition, the atom flux is expressed as

$$J_A^V = -\left(\frac{D_A^V}{C_A} - \frac{G_{\text{Wind}}^A D_B^V}{C_B}\right) C_V \nabla C_A + \left(\frac{D_A^V}{C_A} - \frac{G_{\text{Wind}}^A D_B^V}{C_B}\right) \nabla C_V \tag{S8}$$

$$J_A^I = -\frac{D_A^I}{C_A} C_I \nabla C_A - \frac{D_A^I}{C_A} \nabla C_I \tag{S8}$$

$$J_A = J_A^V + J_A^I \tag{S8}$$

$$J_V = -\left(J_A^V + J_B^V\right) \tag{S8}$$

$$J_I = J_A^I + J_B^I \tag{S8}$$

$D$ is the diffusion coefficient defined by $D_0 \exp(-E_m/kT)$ where $D_0$ is the pre-exponential factor and $E_m$ is the migration energy. $G_{\text{Wind}}^A$ is the vacancy wind factor. The recombination coefficient $R$ is given by

$$R = 4\pi d_{rec} \frac{D_V + D_I}{\Omega_a} \tag{S8}$$

where $d_{rec}$ is the recombination distance.

The equations are numerically solved by implementing SUNDIALS (SUite of Nonlinear and DIfferential/ALgebraic Equation Solvers) [25]. The initial condition of each concentration is determined by the nominal composition in T91 or defect formation energy ($E_f$). The size of 1D computational supercell is determined by the dislocation density by $(\rho_d)^{-0.5}$ = 40 nm with symmetry boundary conditions imposed at the two boundaries. The fitting with experiment data [19] was done by adjusting the pre-exponential factors to find the consistent magnitude of RIS integrated through 2 nm from the dislocation core. Table S5 lists the physical parameters of the RIS model. It is noted that some pre-exponential factors are not available in literature, so in this study we consider them as fitting parameters with reasonable values that were determined by comparing with the experimental RIS measurements at dislocations by Jiao and Was [19], in which the RIS at 7 dpa was measured. Due to limited data available for fitting parameters we assume that the pre-exponential for interstitial diffusion for Ni and Si are the same, yielding only two parameters to be fit, as shown in Table S5. It is noted that because the RIS calculation here is quite simple the fit values are likely not transferable to significantly different temperatures or fluxes. More accurate modeling requires full evaluation of interstitial and vacancy Onsager coefficients as a function of composition, and experimental comparisons over a wide range of conditions. However, the RIS modeling presented here provides approximate evaluation of dose-dependent solute enrichment at different time scales.

Table S5. Physical parameters used in the RIS model. * The variables marked with a *, which include the pre-exponential factors of interstitial diffusivities of Ni and Si and the pre-exponential factor of vacancy diffusivity of Si are determined by fitting with the experimental RIS measurement at 7 dpa from Ref. [7]

| Parameter | Symbol | Value |
|---|---|---|
| Pre-exponential factor of Fe interstitial diffusivity | $D_{0,Fe}^{Int}$ | $6.59 \times 10^{-7}$ m$^2$ s$^{-1}$ [24] |
| *Pre-exponential factor of Ni interstitial diffusivity | $D_{0,Ni}^{Int}$ | $1.25 \times 10^{-7}$ m$^2$ s$^{-1}$ |
| *Pre-exponential factor of Si interstitial diffusivity | $D_{0,Si}^{Int}$ | $0.92 \times 10^{-7}$ m$^2$ s$^{-1}$ |
| Pre-exponential factor of Fe vacancy diffusivity | $D_{0,Fe}^{Vac}$ | $1.02 \times 10^{-4}$ m$^2$ s$^{-1}$ [26] |
| Pre-exponential factor of Ni vacancy diffusivity | $D_{0,Ni}^{Vac}$ | $2.3 \times 10^{-4}$ m$^2$ s$^{-1}$ [27] |
| *Pre-exponential factor of Si vacancy diffusivity | $D_{0,Si}^{Vac}$ | $1.7 \times 10^{-4}$ m$^2$ s$^{-1}$ |
| Migration energy for Fe interstitial diffusivity | $E_{a,Fe}^{Int}$ | 0.36 eV [24] |
| Migration energy for Ni interstitial diffusivity | $E_{a,Ni}^{Int}$ | 0.45 eV [28] |
| Migration energy for Si interstitial diffusivity | $E_{a,Si}^{Int}$ | 0.52 eV [28] |
| Migration energy for Fe vacancy diffusivity | $E_{a,Fe}^{Vac}$ | 0.55 eV [29] |
| Migration energy for Ni vacancy diffusivity | $E_{a,Ni}^{Vac}$ | 0.50 eV [27] |
| Migration energy for Si vacancy diffusivity | $E_{a,Si}^{Vac}$ | 0.51 eV [30] |
| Vacancy wind factor for Ni in Fe | $G_{Wind}^{Ni}$ | -1.6 [30] |
| Vacancy wind factor for Si in Fe | $G_{Wind}^{Si}$ | -1.8 [30] |
| Vacancy formation energy | $E_f^{Vac}$ | 2.00 eV [29] |
| Interstitial formation energy | $E_f^{Int}$ | 3.64 eV [24] |
| Cascade efficiency | $\xi$ | 0.9 [12] |
| Recombination distance | $d_{rec}$ | 5.7×10$^{-10}$ m [11] |

We note that the RIS model is extremely approximate, and really should be considered no more than a physics-based interpolation of segregation as a function of fluence between zero in the unirradiated condition up to the local enrichment concentrations at dislocations in T91 at 7 dpa measured by Jiao and Was [19]. The model therefore cannot be expected to be accurate for irradiation conditions that are different from those used in Jiao and Was [19], such as variations in temperature, flux, fluence, irradiation type and Mn, Ni, Si alloy solute contents. Further, the method used to calculate the point defect concentration in the RIS model is based on a standard lattice recombination mechanism. This differs somewhat from the defect concentrations in the RED model that treats recombination at solute trapped vacancies [11]. This is not an issue in the present proton irradiation case, but a more accurate and self-consistent RIS model will be developed in future research for more general applications. Future work will also add other new physics to standard RIS models, such as consideration of enriched solute interactions and co-segregation. However, these improvements are beyond the scope of this work, which is focused on MNSPs in T91 and not on developing highly accurate RIS models.

**S4. Summary of fitted values and their implications**

The physical parameters used in the present study are listed in Table S2-S4. The vacancy cluster recombination radius, vacancy cluster annealing time, and pre-exponential factors of solute interstitial diffusion and that of vacancy diffusivity of Si have not been evaluated by experiments and simulations previously, so in the present modeling of precipitate evolution, we treated those parameters by applying physical and reasonable values in an attempt to generate simulation results in quantitative agreement with experiments. For the pre-exponential factors of Ni and Si interstitial diffusion in bcc ferrite, we assumed the values to be at the order of $10^{-7}$ m$^2$/s, which is a reasonable magnitude for interstitial diffusion. The dislocation density in irradiated T91 has been reported in the range of $10^{14}$ to $10^{15}$ m$^{-2}$ [31]. A value of $6.25 \times 10^{14}$ m$^{-2}$ was chosen in this study according to the measurement by Penisten [13]. In calculating the heterogeneous nucleation sites on dislocation lines, we also impose a distance of 5 nm from any evolving cluster to avoid precipitate volume overlap. Although our value of 5 nm is a reasonable lower bound considering the size of the precipitate, the model can be improved by the evaluation of better constrained fitting parameters.

## S5. APT reconstruction showing nucleation of Cu-rich precipitates on dislocations [32]

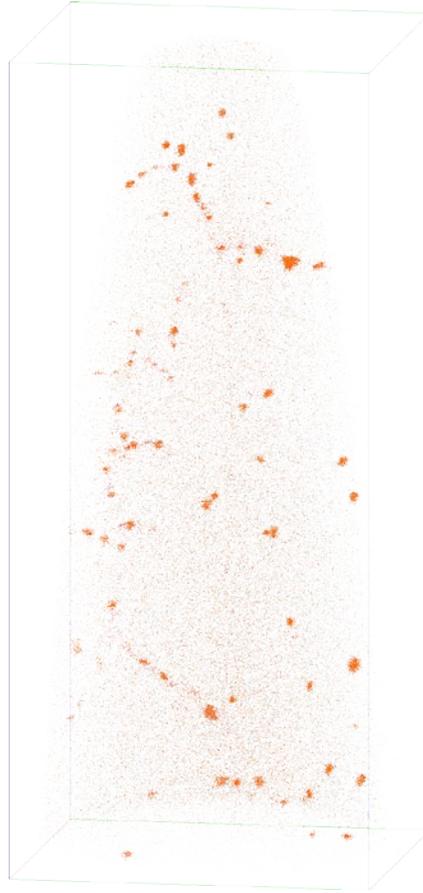

Figure S1. APT reconstruction of the neutron-irradiated LI alloy (see composition in Table S6) showing the preferential precipitation of Cu-rich particles on dislocations [32]. The sample was irradiated under the condition at 310 °C at the flux $3.4\times10^{11}$ n cm$^{-2}$ s$^{-1}$. The fluence is $1.6\times10^{19}$ n cm$^{-2}$. The measured number density and mean radius of Cu-rich precipitates are $8.9\times10^{22}$ m$^{-3}$ and 1.76 nm, respectively.

Table S6. Chemical composition of LI in wt% [32]

|      | C   | Si   | Mn   | P      | Ni   | Cu  | Mo   |
|------|-----|------|------|--------|------|-----|------|
| wt.% | 0.2 | 0.24 | 1.37 | <0.005 | 0.74 | 0.2 | 0.55 |

## S6. Exploration of precipitation kinetics at different temperatures and dpa rates

We utilize the cluster dynamics model integrated with the abovementioned RED and RIS calculations to explore the effects of temperature and dpa rate on the precipitation kinetics of the MNSP at dislocations. Note that due to the limited physics in this model and fitting of multiple parameters to very limited data these calculations for temperature and flux outside the experimental conditions used in fitting are useful only for qualitative guidance and should not be taken to provide quantitative predictions. Figure S2 shows the evolution of number density, mean radius and volume fraction at 400, 350 and 300 °C under the dose rate of $1\times10^{-5}$ dpa/s. The result shows that decreasing temperature slows down the overall nucleation kinetics due to the lower diffusion and longer time for solute enrichment at dislocations. At all temperatures the number density reached to the magnitude of $10^{23}$ m$^{-3}$, which is determined by the dislocation density. Nevertheless, lower temperature provides larger chemical driving force of precipitation, which corresponds to the larger size and volume fraction after higher dose of irradiation. The volume fraction at 300 °C after 100 dpa is able to reach 0.33%, which is larger than that of 0.23% at 400 °C.

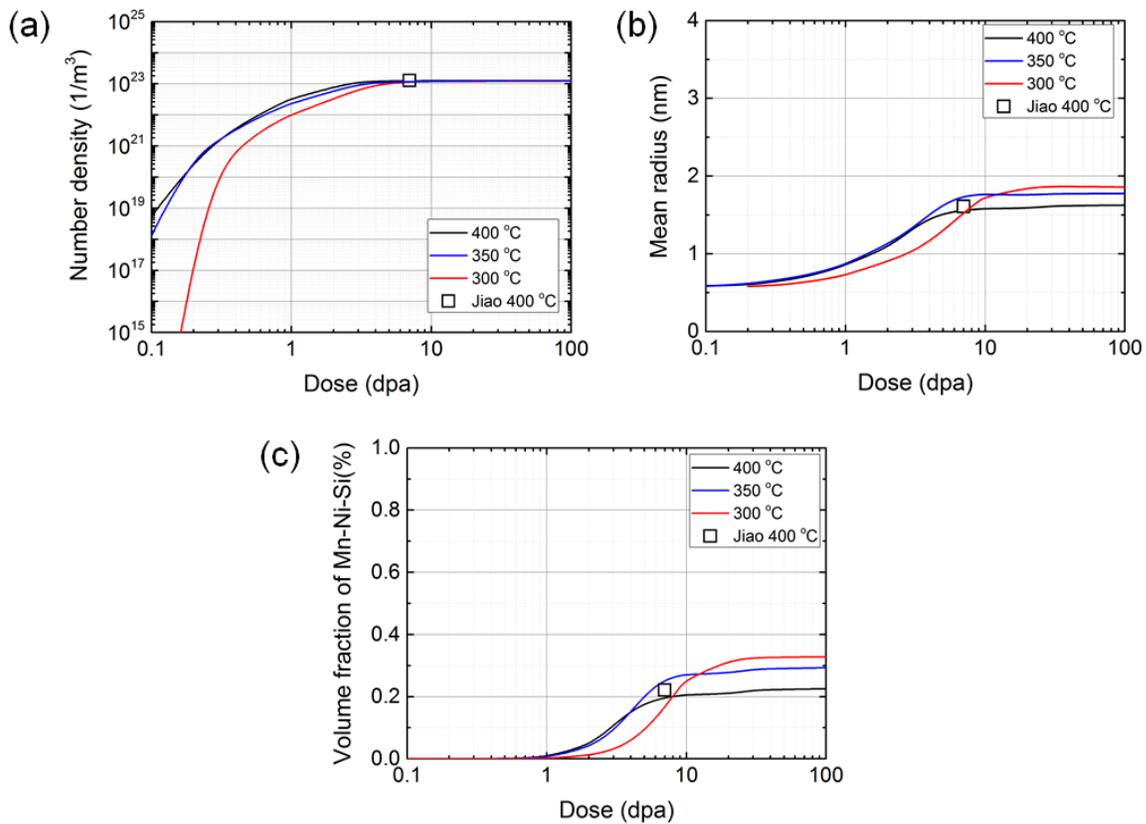

Figure S2. Calculation result of the cluster dynamics model showing the (a) number density, (b) mean radius, and (c) volume fraction of MNSPs as a function of irradiation dose (dpa). The solid lines show the results at 300, 350, and 400 °C under the dose rate of $1\times10^{-5}$ dpa/s. The symbols show the values reported by the experiment [5] at 400 °C.

Figure S3 shows evolution of number density, mean radius and volume fraction of MNSP at 400 °C under the dose rate of $10^{-5}$ and $10^{-7}$ dpa/s, where the latter is the typical magnitude of dpa rate under fast neutron irradiation. The comparison shows that under lower-flux irradiation of $10^{-7}$ dpa/s, the onset of nucleation is reached at a lower dose compared to irradiation at $10^{-5}$ dpa/s. Both the size and number density during lower-flux irradiation saturates at a lower dpa than that under higher-flux irradiation. It is noted that the saturated volume fraction and size at $10^{-7}$ dpa/s are both smaller than that under $10^{-5}$ dpa/s. The difference is because of the less RIS level under low-flux irradiation, which cannot produce larger defect concentration gradient around dislocations to drag solute atoms.

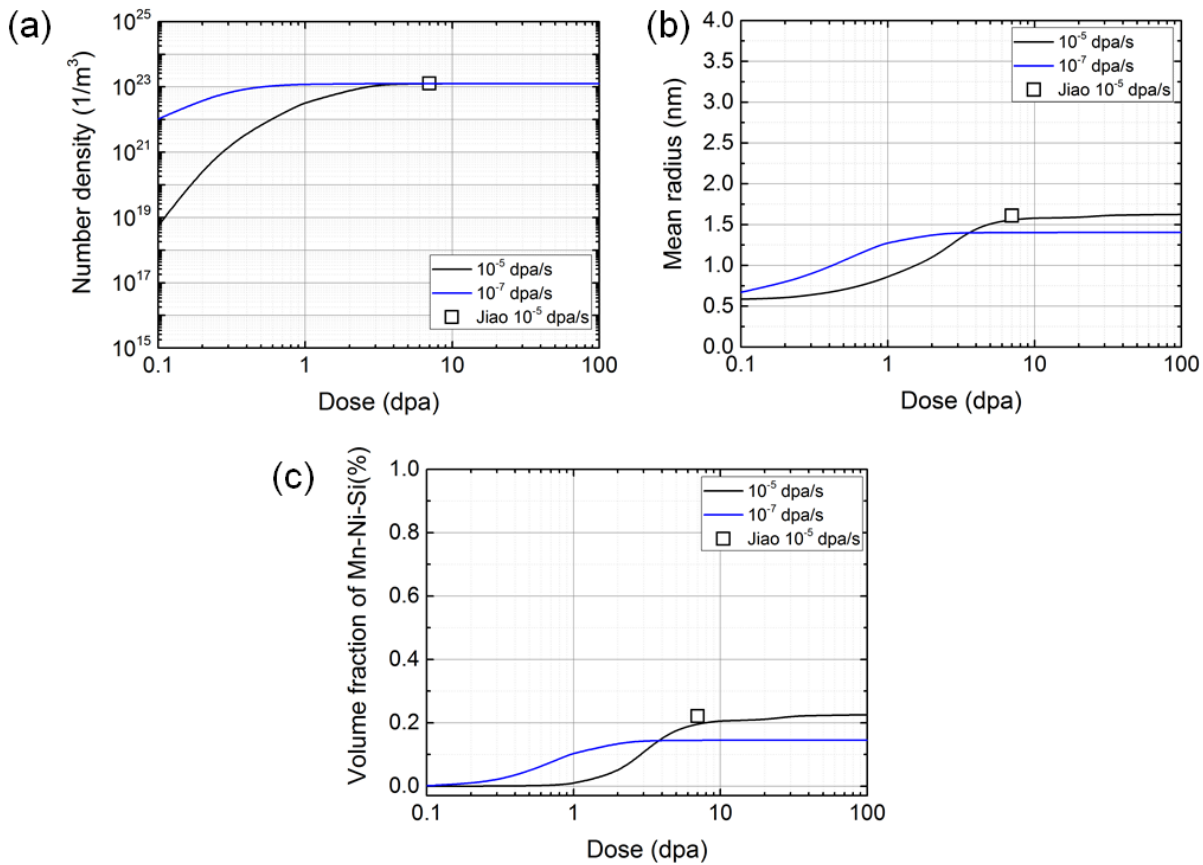

Figure S3. Calculation result of the cluster dynamics model showing the (a) number density, (b) mean radius, and (c) volume fraction of MNSPs as a function of irradiation dose (dpa). The solid lines show the results under the dose rates of $10^{-5}$ and $10^{-7}$ dpa/s at 400 °C. The symbols show the values reported by the experiment [5] at 400 °C.

## S7. Effect of *A* factor on the Mn-Si-Ni precipitation kinetics in proton irradiated T91

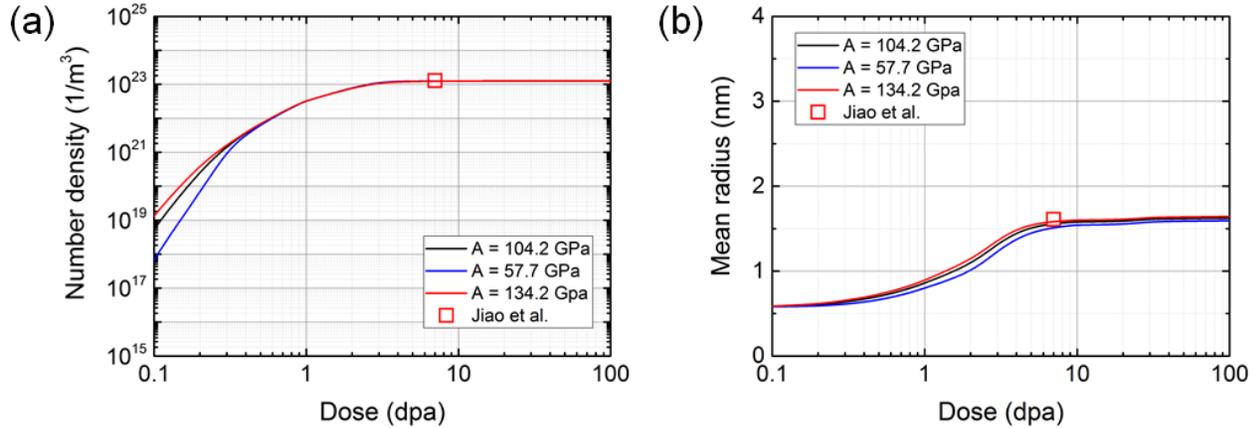

Figure S4. Simulation results showing the effect of *A* factor (Eq. (S13) and (S14)) on the evolution of (a) number density and (b) mean radius of MNSPs at 400 °C and a dose rate of $10^{-5}$ dpa/s. The symbols show the values reported by the experiment [5] at 400 °C.